\documentclass[twocolumn]{IEEEtran}
\usepackage{amsmath,amssymb,mathrsfs,epsfig}

\begin{document}

\title{Numerical Aperture of\\ Single-Mode Photonic Crystal Fibers}

\author{Niels Asger Mortensen, Jacob Riis Folkenberg, Peter M.~W. Skovgaard, and Jes Broeng\thanks{The authors are with the company {\it Crystal Fibre A/S}, Blokken 84, DK-3460 Birker\o d, Denmark; http://www.crystal-fibre.com.}}

\markboth{{\it S\lowercase{ubmitted to}} IEEE Photonics Technology Letters, 2002}
{Mortensen {\it et al.}: Numerical Aperture of Single-mode Photonic Crystal Fibers}

\maketitle

\begin{abstract}
We consider the problem of radiation into free space from the end-facet of a single-mode photonic crystal fiber (PCF). We calculate the numerical aperture ${\rm NA}=\sin\theta$ from the half-divergence angle $\theta \sim \tan^{-1}(\lambda/\pi w)$ with $\pi w^2$ being the effective area of the mode in the PCF. For the fiber first presented by Knight {\it et al.} we find a numerical aperture ${\rm NA} \sim 0.07$ which compares to standard fiber technology. We also study the effect of different hole sizes and demonstrate that the PCF technology provides a large freedom for NA-engineering. Comparing to experiments we find good agreement.
\end{abstract}


\section{Introduction}

\PARstart{P}{hotonic crystal fibers} (PCF) constitute a completely new class of optical fibers consisting of pure silica with air-holes distributed in the cladding. Among many remarkable properties \cite{opticsexpress} PCFs are believed to have a potential for high-numerical aperture (NA) applications. Here we report a calculation of the NA for the class of PCFs first fabricated by Knight {\it et al.} \cite{knight1996,knight1997errata}. For this particular fiber we find a numerical aperture up to ${\rm NA} \sim 0.07$. We also demonstrate how the NA may be controlled by the hole size for a given pitch and wavelength.

\begin{figure}[h!]
\begin{center}
\epsfig{file=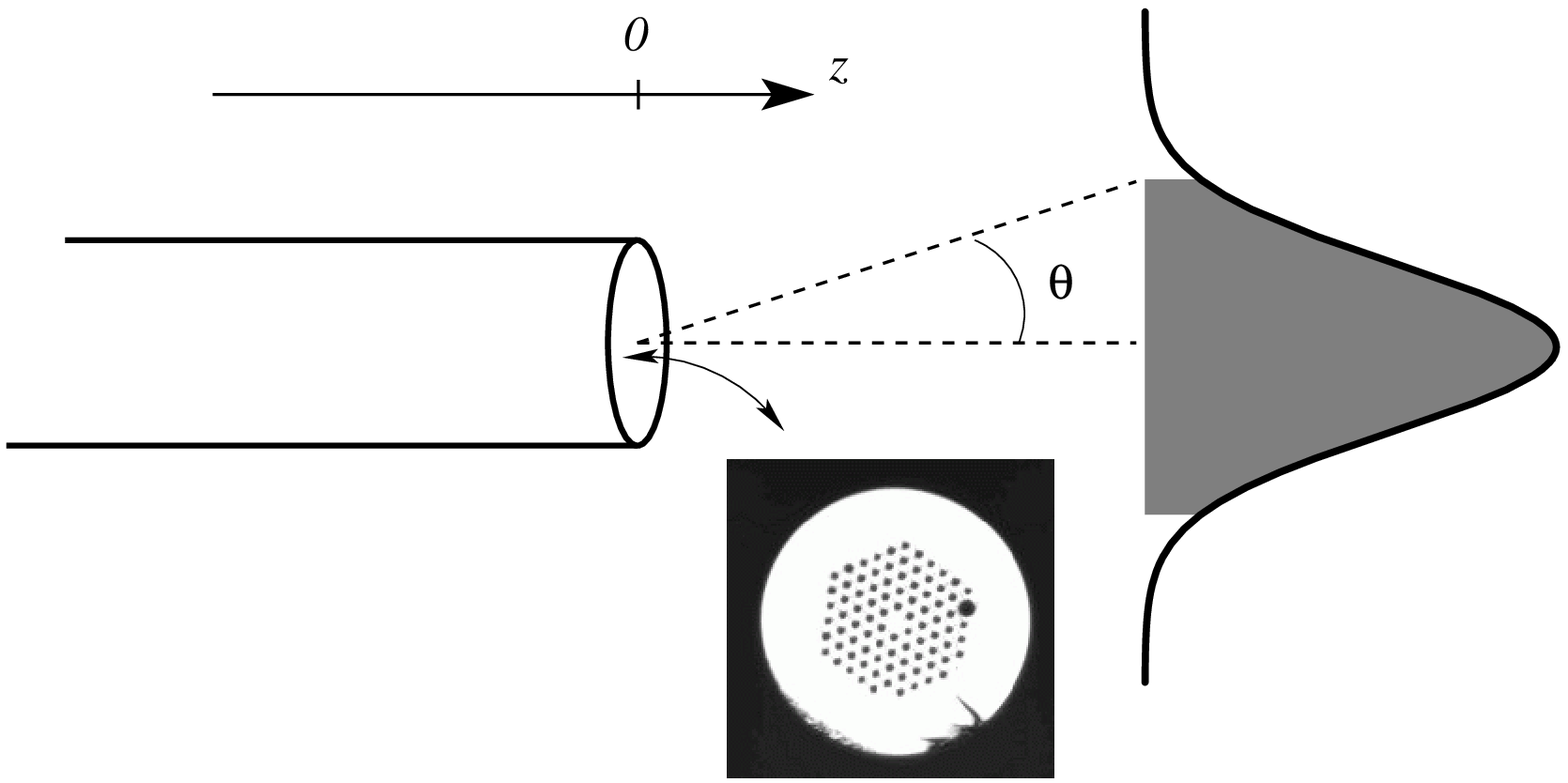, width=0.49\textwidth,clip}
\end{center}
\caption{Coupling of light from end-facet of fiber ($z=0$) into free space. The insert shows a micrograph of the end-facet of a PCF.}
\label{fig_geometry}
\end{figure}

The paper is organized as follows: First we consider the problem of radiation into free space from the end-facet of a single-mode optical fiber with a mode approximated by a Gaussian of width $w$. Solving the scattering problem at the end-facet of the fiber exactly we check the range of validity of the text-book result $\theta \simeq \tan^{-1}(\lambda/\pi w)$ (see {\it e.g.} Ref.~\cite{ghatak1998}) where $A_{\rm eff}=\pi w^2$ is the effective area. For $w\ll \lambda$ we find deviations whereas nice agreement is found for $w>\lambda$. We then turn to the application of the PCF of Knight {\it et al.} \cite{knight1996} which belongs to the latter regime with $w> \lambda$. Finally, we compare our calculations to experiments.

\section{Numerical aperture in the Gaussian approximation}

The numerical aperture ${\rm NA}=\sin\theta$ (see Fig.~\ref{fig_geometry}) may be defined in various ways, but often one defines it in the far-field limit ($z\rightarrow \infty$) from the half-divergence angle $\theta_\nu$ between the $z$-axis and the $\nu$-intensity point $r_\nu(z)$, {\it i.e.}

\begin{equation}
\tan\theta_\nu = \lim_{z\rightarrow \infty}\frac{r_\nu(z)}{z}, 
\end{equation}
with $r_\nu(z)$ determined from
\begin{equation}
\frac{\big|\Psi_>(z,r_\perp=r_\nu)\big|^2}{\big|\Psi_>(z,r_\perp=0)\big|^2}=\nu.
\end{equation}
 For a Gaussian field $\Psi$ of width $w$ one has the standard approximate expression for $\nu=1/e^2\simeq 13.5\%$ \cite{ghatak1998}

\begin{equation}\label{ghatak}
\tan\theta_{1/e^2}\simeq \frac{2}{kw} = \frac{\lambda}{\pi w}.
\end{equation}
For the $\nu=5\%$ intensity point
\begin{equation}\label{ghatak_5}
\tan\theta_{5\%}=\sqrt{\frac{\ln 20}{2}}\times\tan\theta_{1/e^2}
\end{equation}
which is often the one used experimentally. Eqs.~(\ref{ghatak},\ref{ghatak_5}) are valid for $kw\gg 1$, but in order to check the validity in the limit with $kw$ of order unity we solve the scattering problem at the end-facet of the fiber exactly. In the fiber ($z<0$) the field is of the form 
\begin{equation}
\Psi_<(r)\propto \psi(r_\perp)\big( e^{i\beta(\omega) z}+{\mathscr R} e^{-i\beta(\omega) z}\big),\;z<0
\end{equation}
where the transverse field is approximated by a Gaussian
\begin{equation}
\psi(r_\perp)\propto e^{-(r_\perp/w)^2},
\end{equation}
which has an effective area $A_{\rm eff}=\pi w^2$ at frequency $\omega$. At the end-facet of the fiber ($z=0$) the field couples to the free-space solution
\begin{equation}\label{Psi>}
\Psi_>(r)\propto \int d{\boldsymbol k}_\perp {\mathscr T}({\boldsymbol k}_\perp) e^{i {\boldsymbol k}_\perp \cdot {\boldsymbol r}_\perp}e^{ik_\parallel z} ,\;z>0
\end{equation}
which is a linear combination of plane waves with $\omega=ck=c(2\pi/\lambda)$ and ${\boldsymbol k}={\boldsymbol k}_\perp + {\boldsymbol k}_\parallel$.

In order to solve the elastic scattering problem, $\Delta\omega=\omega(\beta)-\omega(k)=0$, we apply appropriate boundary conditions at the end-facet of the fiber; continuity of $\Psi$ and $\partial 
\Psi /\partial z$. At $z=0$ we thus get two equations determining the reflection amplitude ${\mathscr R}$ and the transmission amplitude ${\mathscr T}$. Eliminating ${\mathscr R}$ and substituting the resulting ${\mathscr T}$ into Eq.~(\ref{Psi>}) we get

\begin{eqnarray}\label{Psi>_final}
\Psi_>(r)&\propto&2\pi k^2\int_0^\infty d\chi\,\chi \frac{2n_{\rm eff}}{\sqrt{1-\chi^2}+n_{\rm eff}}\nonumber\\
&&\qquad\times e^{-(\chi kw/2)^2}J_0(\chi kr_\perp)e^{i \sqrt{1-\chi^2} kz}.
\end{eqnarray}
Here,  $\chi=k_\perp/k$, $J_0$ is the Bessel function of the first kind of order $0$, and $n_{\rm eff} =\beta/k$ is the effective mode-index. Eq.~(\ref{Psi>_final}) is the exact solution to the scattering problem and in contrast to many approximate text-book expressions (see {\it e.g.} Refs.~\cite{ghatak1998}) we have here treated the scattering problem correctly including the small, but finite, backscattering in the fiber. Thus, we take into account the possible filtering in transmitted $k_\perp$ at the fiber end-facet. The solution has similarities with the Hankel transform usually employed in the far-field inversion integral technique, see {\it e.g.} \cite{anderson1983}. Numerically we have found that Eq.~(\ref{Psi>_final}) gives a close-to-Gaussian field in the far-field limit.

\begin{figure}[h!]
\begin{center}
\epsfig{file=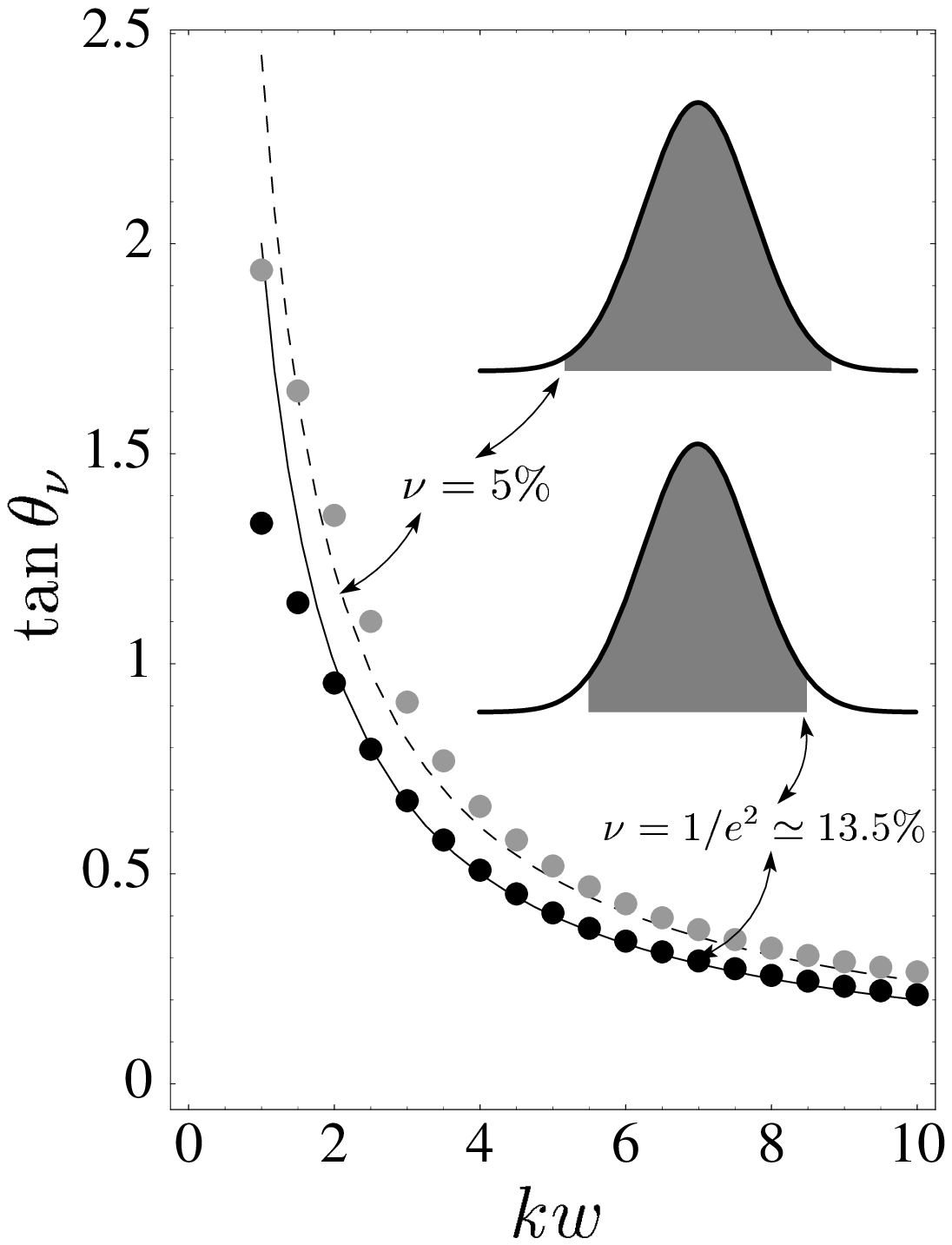, width=0.45\textwidth,clip}
\end{center}
\caption{Plot of $\tan\theta$ as a function of the dimensionless parameter $kw$. The points are the results of a numerical exact calculation from Eq.~(\ref{Psi>_final}) for a mode with effective index, $n_{\rm eff}=\beta/k=1.444$. The full and dashed lines show the approximations Eqs.~(\ref{ghatak},\ref{ghatak_5}), respectively.}
\label{fig_NA_gauss}
\end{figure}

In Fig.~\ref{fig_NA_gauss} we compare the two approximate solutions Eqs.~(\ref{ghatak},\ref{ghatak_5}) to a numerically exact calculation of $\tan\theta_\nu$ from Eq.~(\ref{Psi>_final}). The calculation is performed for the realistic situation with $n_{\rm eff}=\beta/k=1.444$ corresponding to a silica-based fiber. For $kw\sim 1$ the deviations increase because of the small, but finite, backscattering at the end-facet of the fiber. For $kw$ somewhat larger than unity a very nice agreement is found. A typical all-silica fiber like the Corning SMF28 has $kw >10$.

\begin{figure}[h!]
\begin{center}
\epsfig{file=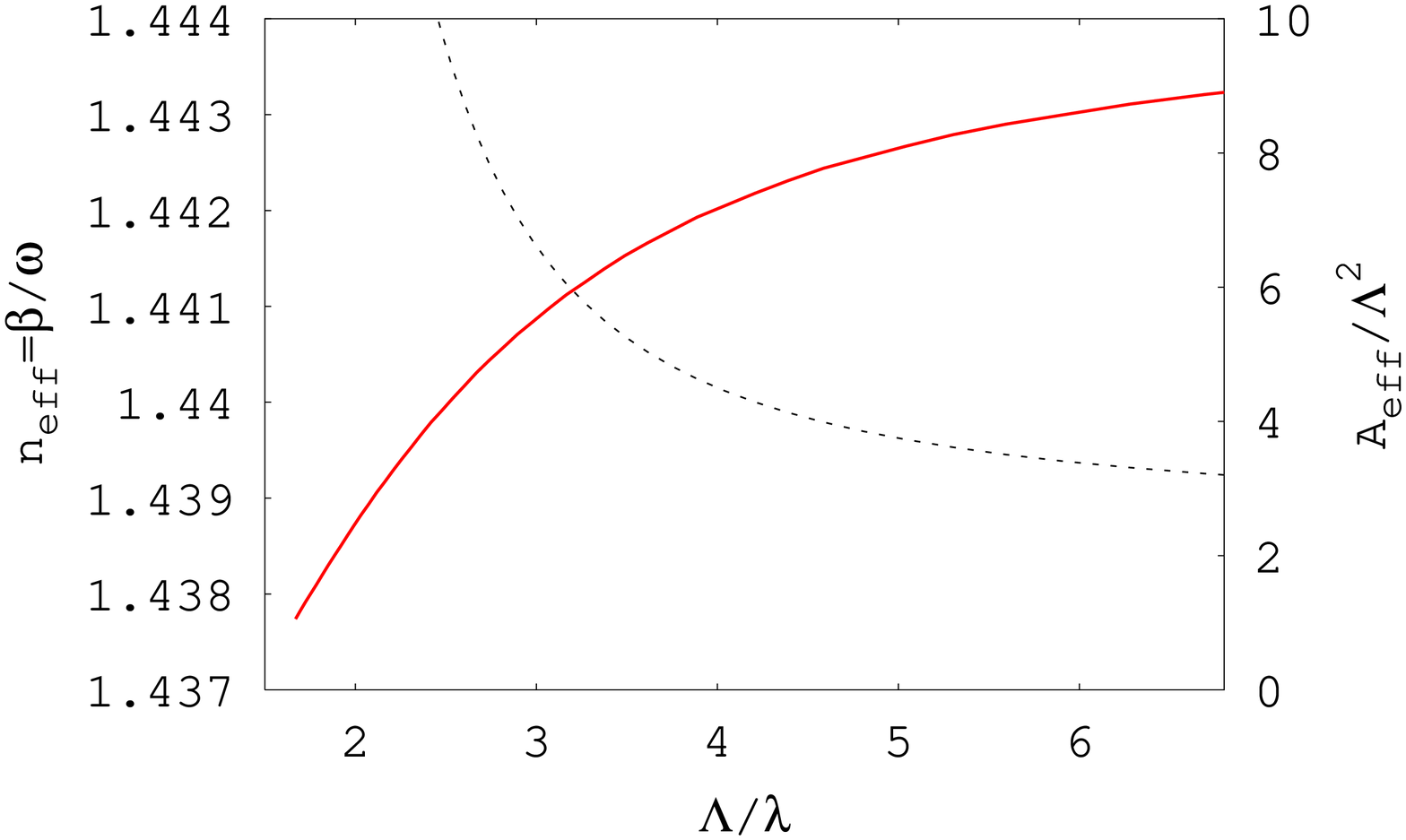, width=0.48\textwidth,clip}
\end{center}
\caption{Effective mode-index (solid line, left axis) and effective area (dashed line, right axis) of a PCF with $d/\Lambda=0.15$.}
\label{fig_index}
\end{figure}

\section{Application to photonic crystal fibers}
 We consider the class first studied in Ref.~\cite{knight1996} which consists of pure silica with a cladding with air-holes of diameter $d$ arranged in a triangular lattice with pitch $\Lambda$. For a review of the operation of this class of PCFs we refer to Ref.~\cite{broeng1999}.

In applying Eq.~(\ref{ghatak}) to PCFs we calculate $w$ from the effective area $A_{\rm eff}=\pi w^2$ given by~\cite{agrawal} 

\begin{equation}
A_{\rm eff}= \frac{\big[\int d{\boldsymbol r} \big|{\boldsymbol H}({\boldsymbol r},z)\big|^2\big]^2}{\int d{\boldsymbol r} \big|{\boldsymbol H}({\boldsymbol r},z)\big|^4}.
\end{equation}
Indeed we find that the corresponding Gaussian of width $w$ accounts well for the overall spatial dependence of the field. Of course we thereby neglect the satellite spots seen in the far-field \cite{knight1996}, but because of their low intensity they only give a minor contribution to the NA~\cite{mortensen_unpublished}.  

For the field $\boldsymbol H$ of the PCF, fully-vectorial eigenmodes of Maxwell's equations with periodic boundary conditions are computed in a planewave basis \cite{johnson2000}.

\begin{figure}[h!]
\begin{center}
\epsfig{file=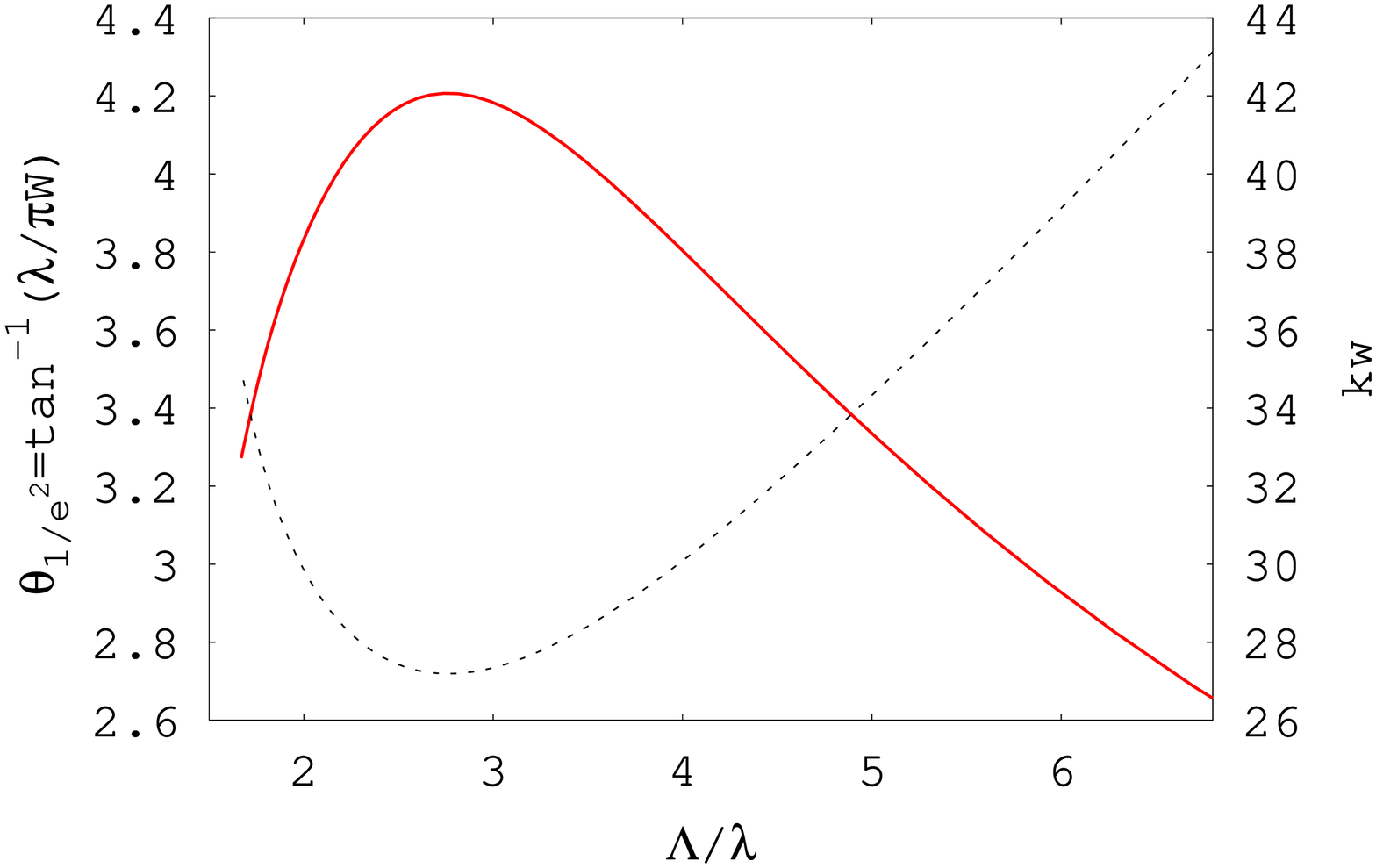, width=0.48\textwidth,clip}
\end{center}
\caption{Half-divergence angle (solid line, left axis) and $kw$ (dashed line, right axis) of a PCF with $d/\Lambda=0.15$.}
\label{fig_NA}
\end{figure}
Figure \ref{fig_index} illustrates the effective mode-index and effective area as a function of wavelength for a PCF with $d/\Lambda=0.15$. The first PCF fabricated by Knight {\it et al.} \cite{knight1996} of this kind had a pitch of $\Lambda = 2.3\,{\rm \mu m}$ and was found to be single-mode in the range $\Lambda/\lambda$ between $1.5$ and $6.8$. In Fig.~\ref{fig_NA} we show the corresponding half-divergence angle. We have also shown the value of the dimensionless parameter $kw$ (dashed line, right axis); the magnitude justifies the application of the approximate result in Eq.~(\ref{ghatak}) to PCFs. We note that for non-linear PCFs \cite{knight2000} the value of $kw$ will approach the regime where deviations from Eq.~(\ref{ghatak}) arise.

\begin{figure}[h!]
\begin{center}
\epsfig{file=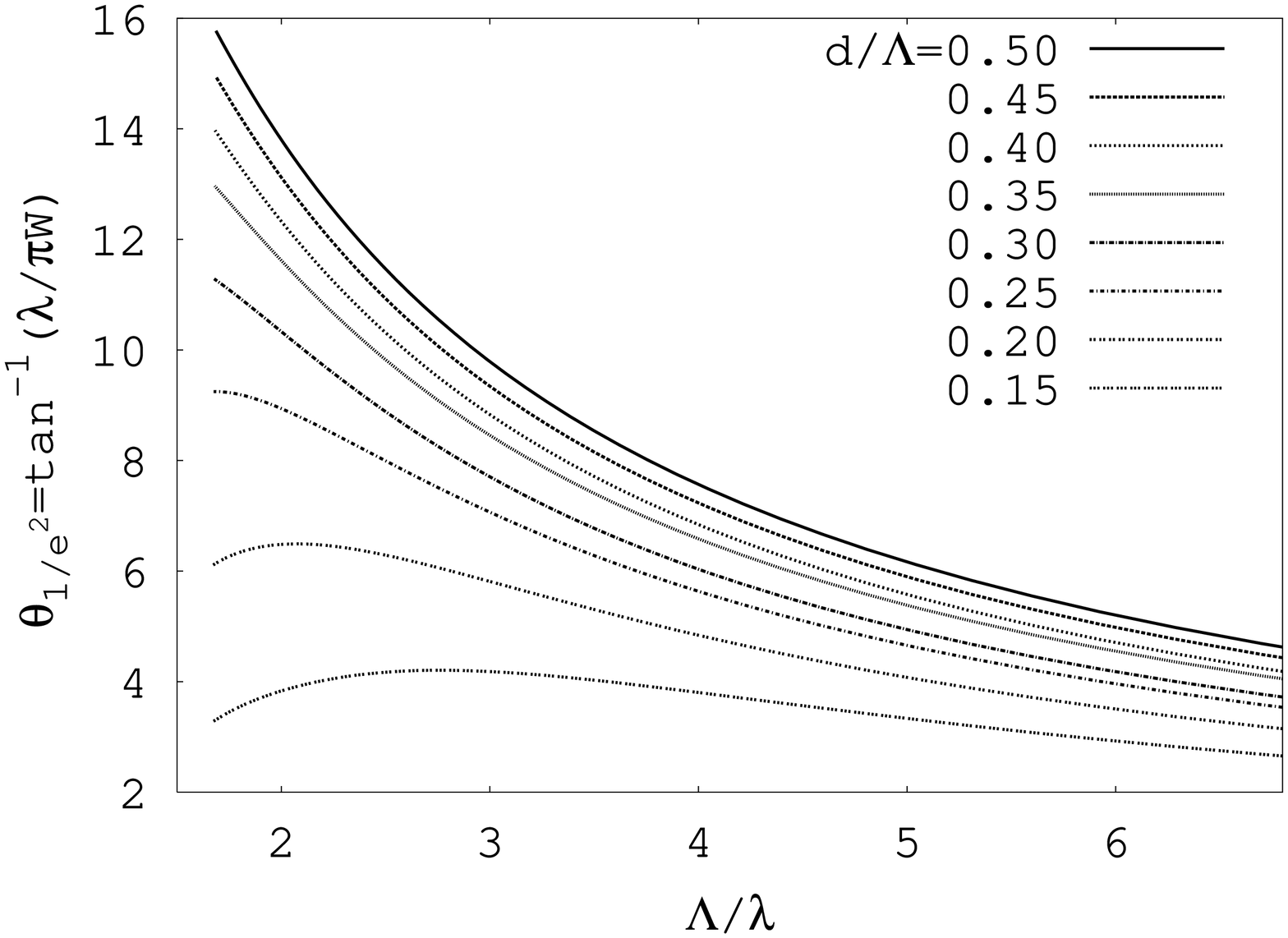, width=0.48\textwidth,clip}
\end{center}
\caption{Half-divergence angle of a PCF for different hole sizes.}
\label{fig_NA_d}
\end{figure}

In Fig.~\ref{fig_NA_d} we show the half-divergence angle for different hole sizes where the fiber is endlessly single mode \cite{birks1997}. For small hole sizes $d/\Lambda$ we note that in practice the operation is limited by a significant confinement loss for long wavelengths where the effective area increases~\cite{white2001}. In Fig.~\ref{fig_NA_d} this can be seen as a bending-down of $\theta$ for small $\Lambda/\lambda$. In general the NA increases for increasing hole size and fixed pitch and wavelength. By adjusting the pitch $\Lambda$ and the hole size $d$ this demonstrates a high freedom in designing a fiber with a certain NA at a specified wavelength. 

\begin{figure}[h!]
\begin{center}
\epsfig{file=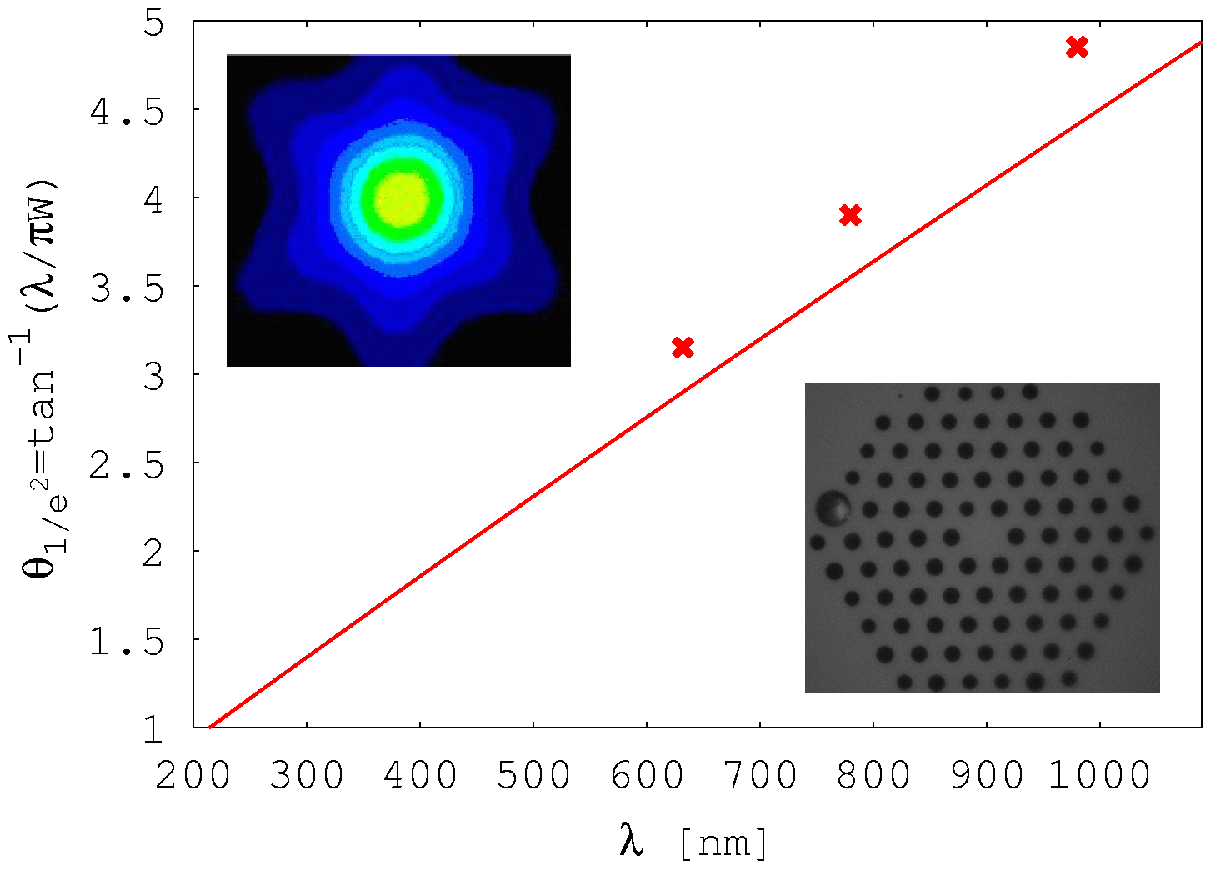, width=0.48\textwidth,clip}
\end{center}
\caption{Half-divergence angle of a PCF with $d/\Lambda\sim 0.53$ and $\Lambda\simeq 7.2\,{\rm  \mu}m$. The solid line is a calculation based on the ideal structure and the data points are measurements at $\lambda = 632\,{\rm nm}$, $780\,{\rm nm}$, and $980\,{\rm nm}$ of the fiber shown in the lower right insert. The upper left insert shows a near-field image at $\lambda = 632\,{\rm nm}$.}
\label{fig_NA_experiment}
\end{figure}

In order to verify our calculations experimentally a PCF with $d/\Lambda\sim 0.53$ and $\Lambda\simeq 7.2\,{\rm  \mu}m$ has been fabricated. In Fig.~\ref{fig_NA_experiment} we compare our calculations to a measurement of the NA at the wavelength $\lambda = 632\,{\rm nm}$. As seen the calculation agrees well with the measured value.

\section{Conclusion}

We have studied the numerical aperture (NA) of photonic crystal fibers (PCF). The calculations is based on the approximate ``standard'' result $\theta \simeq \tan^{-1}(\lambda/\pi w)$ which we have found to be valid in the regime relevant to PCFs. As an example we have applied it to the fiber first fabricated by Knight {\it et al.} \cite{knight1996}. By studying the effect of different hole sizes we have demonstrated that the PCF technology have a strong potential for NA-engineering in the single-mode regime.


\begin{thebibliography}{10}

\bibitem{opticsexpress}
For recent special issues on photonic crystal fibers, see {\it Opt.
  Express}, vol. {\bf 9}, pp.~674--779, 2001; J. Opt. A: Pure Appl. Opt., vol.
  {\bf 3}, pp.~S103--S207, 2001.

\bibitem{knight1996}
J.~C. Knight, T.~A. Birks, P.~St.~J. Russell, and D.~M. Atkin,
\newblock ``All-silica single-mode optical fiber with photonic crystal
  cladding,''
\newblock {\em Opt. Lett.}, vol. 21, pp. 1547--1549, 1996.

\bibitem{knight1997errata}
J.~C. Knight, T.~A. Birks, P.~St.~J. Russell, and D.~M. Atkin,
\newblock ``All-silica single-mode optical fiber with photonic crystal
  cladding: errata,''
\newblock {\em Opt. Lett.}, vol. 22, pp. 484--485, 1997.

\bibitem{ghatak1998}
A.~K. Ghatak and K.~Thyagarajan,
\newblock {\em Introduction to Fiber Optics},
\newblock Cambridge University Press, Cambridge, 1998.

\bibitem{anderson1983}
W.~T. Anderson and D.~L. Philen,
\newblock ``Spot size measurements for single-mode fibers -- a comparison of
  four techniques,''
\newblock {\em J. Lightwave Technol.}, vol. 1, pp. 20--26, 1983.

\bibitem{broeng1999}
J.~Broeng, D.~Mogilevstev, S.~E. Barkou, and A.~Bjarklev,
\newblock ``Photonic crystal fibers: A new class of optical waveguides,''
\newblock {\em Opt. Fiber Technol.}, vol. 5, pp. 305--330, 1999.

\bibitem{agrawal}
G.~P. Agrawal,
\newblock {\em Nonlinear Fiber Optics},
\newblock Academic Press, San Diego, 2001.

\bibitem{mortensen_unpublished}
N. A. Mortensen and J. R. Folkenberg, \newblock ``Near-field to far-field transition of photonic crystal fibers: symmetries and interference phenomina,'' {\em Opt. Express}, vol. 10, pp. 475-481, 2002.

\bibitem{johnson2000}
S.~G. Johnson and J.~D. Joannopoulos,
\newblock ``Block-iterative frequency-domain methods for \uppercase{M}axwell's
  equations in a planewave basis,''
\newblock {\em Opt. Express}, vol. 8, pp. 173--190, 2001.

\bibitem{knight2000}
J.~C. Knight, J.~Arriaga, T.~A. Birks, A.~Ortigosa-Blanch, W.~J. Wadsworth, and
  P.~St.~J. Russell,
\newblock ``Anomalous dispersion in photonic crystal fiber,''
\newblock {\em IEEE Photonic Tech. L.}, vol. 12, pp. 807--809, 2000.

\bibitem{birks1997}
T.~A. Birks, J.~C. Knight, and P.~St.~J. Russell,
\newblock ``Endlessly single mode photonic crystal fibre,''
\newblock {\em Opt. Lett.}, vol. 22, pp. 961--963, 1997.

\bibitem{white2001}
T.~P. White, R.~C. McPhedran, C.~M. {de Sterke}, L.~C. Botton, and M.~J. Steel,
\newblock ``Confinement losses in microstructured optical fibers,''
\newblock {\em Opt. Lett.}, vol. 26, pp. 1660--1662, 2001.

\end{thebibliography}
\end{document}